\documentclass[usegraphicx,useAMS,usenatbib]{mn2e}

\usepackage{graphicx}
\usepackage{rotating}

\title[Radio-mode AGN feedback in Post-starburst galaxies]{Radio Emission and AGN Feedback in Post-starburst Galaxies}

\author[M.-S. Shin et al.]
{Min-Su Shin,$^{1,2}$\thanks{E-mail:msshin@umich.edu,
strauss@astro.princeton.edu,
rita.tojeiro@port.ac.uk}
Michael~A.~Strauss,$^2$
Rita Tojeiro$^3$\\
$^1$Department of Astronomy, University of Michigan, Ann Arbor, MI 48109, USA\\
$^2$Princeton University Observatory, Peyton Hall, Princeton, NJ 08544-1001, USA\\
$^3$Institute of Cosmology and Gravitation, Dennis Sciama Building, Burnaby Road, Portsmouth, PO1 3FX
}

\date{Released 2010 Xxxxx XX}

\newcommand{\simgt} {\,\hbox{\lower0.6ex\hbox{$\sim$}\llap{\raise0.6ex\hbox{$>$}}}\,}
\newcommand{\simlt} {\,\hbox{\lower0.6ex\hbox{$\sim$}\llap{\raise0.6ex\hbox{$<$}}}\,}
\newcommand{\nodata} { ~$\cdots$~ }

\begin{document}

\date{Accepted ... Received ..; in original form ..}

\maketitle

\begin{abstract}
We investigate radio-mode AGN activity among post-starburst 
galaxies from the Sloan Digital Sky Survey to determine whether 
AGN feedback may be responsible for the cessation of star formation. 
Based on radio morphology and radio-loudness from 
the FIRST and NVSS data, we 
separate objects with radio activity 
due to an AGN from ongoing residual star formation. 
Of 513 SDSS 
galaxies with strong A-star spectra, 12 objects have 21-cm flux density above 
1 mJy. These galaxies do not show optical AGN emission lines. 
Considering that the lifetime of radio emission is much shorter than the typical time-scale of the 
spectroscopic features of post-starburst galaxies, we conclude that the radio-emitting AGN activity in these objects 
was triggered after the end of the recent starburst, and thus cannot be 
an important feedback process to explain the post-starburst phase. The 
radio luminosities show a positive correlation with total galaxy stellar mass, but not with 
the mass of recently formed stars. Thus the mechanical power of AGN feedback 
derived from the radio luminosity is related to old stellar populations dominating the stellar mass, which 
in turn are related to the masses of central supermassive black holes.
\end{abstract}

\begin{keywords}
galaxies: evolution\ -- galaxies: fundamental parameters\ -- 
galaxies: jets\ -- galaxies: stellar content\ -- radio continuum: general
\end{keywords}

\section{Introduction \label{sec:intro}}

Various observational facts and theoretical models 
have suggested that feedback from AGN activity 
has a strong impact on galaxy formation and evolution, especially 
star formation in galaxies \citep[see][for a review]{bland07}. 
The starburst-AGN 
connection has been investigated for a broad range of galaxy and AGN types, 
showing that the mechanism triggering intensive star formation might 
also cause strong fuelling onto a central supermassive black hole \citep[e.g.][]
{smith98,cid01,veilleux01,farrah03,rupke05,lonsdale06,liu09}. Effects of 
AGN activity on the 
interstellar medium and its evolution are traditionally split into two categories: 
the radiative mode, including ionisation, heating, and radiation pressure \citep[e.g.][]{ciotti07} and 
the mechanical mode via nuclear winds and jets \citep[e.g.][]{ciotti09,shin10}, although the 
physics of these feedback process is still uncertain \citep[see][for a review]{begelman04}. 
AGN feedback in many different forms has been highlighted as a possible mechanism to 
explain the shape of the galaxy luminosity function 
\citep[e.g.][]{croton06}, the temperature-luminosity relationship of galaxy clusters 
\citep[e.g.][]{bower01}, the cooling flow problem of galaxy clusters \citep[e.g.][]
{fabian94,vernaleo06}, the colour bimodality of galaxies \citep[e.g.][]{smolvcic09}, 
and other observational phenomena. 
The form of AGN feedback varies among different models, but 
the main idea of this AGN feedback is to regulate 
the supply of potentially star-forming cold gas 
by supplying extra energy to the interstellar and intergalactic medium.

The effects of AGN on star formation can be examined in several different stages of galaxy evolution. 
One way is to understand the stellar populations of AGN host galaxies \citep[see][for a review]
{canalizo06}. For example, 
the host galaxies of low-redshift narrow-line quasars 
have a significant fraction of recently formed (1 $<$ Gyr) stars \citep{kauffmann03a,liu09}. 
\citet{vandenberk06} found similar results for broad-line quasars. 
The mass of molecular gas varies in AGN host galaxies, and is also an indirect measurement 
of their star-formation potential \citep[e.g.][]{papadopoulos08,evans09,wang10}. 
Another way is to investigate galaxies such as ULIRGs, 
hosting both AGNs and intensive star formation simultaneously 
\citep[see][for a review]{sanders96}. Several radio-loud star-forming AGNs also 
show a connection between the suppression of star 
formation and the strength of radio jets \citep[e.g.][]{nesvadba08}.

As a special and rare stage in galaxy evolution, 
post-starburst galaxies, also known as K+A or E+A galaxies because of 
the characteristic features in their spectra, 
have recently undergone an abrupt cessation of active star formation. 
These galaxies  
do not show signs of current active star formation in the form of optical 
emission lines, but their stellar populations are composed both of 
recently formed stars (represented by A-type stars 
with their strong Balmer absorption lines) and of old populations 
\citep[e.g.][]{dressler83,zabludoff96,quintero04}. 
Post-starburst galaxies are located in similar large-scale environments as are 
normal star-forming and quiet galaxies \citep{blake04,hogg06}, even though many post-starburst 
galaxies show dynamical interactions with neighbours \citep{yagi06}. 
Field E+A galaxies 
are quite heterogeneous in terms of their surface brightness distribution, 
velocity dispersion, and luminosity despite their similar 
spectroscopic features \citep{tran04}.

There are several models to explain what causes the intensive star formation and is abrupt truncation 
in post-starburst 
galaxies. External effects on galaxies such as ram pressure stripping might play an important role in 
quickly ceasing star formation \citep{gunn72,dressler84}. 
Galaxy mergers are also thought to 
trigger intensive star formation, and to stop it quickly without requiring feedback from 
a central supermassive black hole \citep{bekki05} on timescales from 
a few hundred Myr to few Gyr \citep[see][for timescales and star formation models]{falkenberg09a}.
Alternatively, the truncation of star formation might be caused by internal processes 
such as AGN feedback effects, 
although we do not know yet what types of AGN feedback effects are most important 
\citep{ciotti07,ciotti09,shin10,ciotti10}. 
More general scenarios can be constructed by combining the internal and external processes 
described above \citep[e.g.][]{springel05,sijacki07,khalatyan08}.

If the sudden cessation of star formation in post-starburst galaxies is related to AGN 
activity, we might expect to see several different forms 
of AGN activity in these objects. 
X-ray emission due to AGN has been reported in some 
post-starburst galaxies 
\citep{dewangan00,georgakakis08,brown09}. Some post-starburst galaxies also show 
AGN emission lines in the optical in addition to strong Balmer absorption lines 
\citep{brotherton99,goto06,yan06,yang06}. 
Radio continuum emission thought be due to radio-mode AGN activity is also detected in some post-starburst galaxies 
\citep[e.g.][]{liu07}. Outflows have been measured via 
MgII absorbers in some luminous post-starburst galaxies, 
with velocities comparable to those due to 
AGN or extreme starbursts \citep{tremonti07}.

\begin{table*}
\begin{minipage}{126mm}
\caption{Radio-emitting post-starburst galaxies \label{tab:list}}
\begin{tabular}{@{}lccccc} \hline \hline
SDSS name & Redshift & $F_{int}^{FIRST}$\footnotemark[1]
& $RMS^{FIRST}$\footnotemark[2]
& $F^{NVSS}$\footnotemark[3]
& $RMS^{NVSS}$\footnotemark[4]\\
& & (mJy) & (mJy) &(mJy) & (mJy) \\ \hline
SDSSJ082254.8+192128 & 0.0626 & 0.70 & 0.15 & \nodata & \nodata \\
SDSSJ084542.7+292932 & 0.1475 & 0.96 & 0.14 & \nodata & \nodata \\
SDSSJ092023.1+394039 & 0.0690 & 7.96 & 0.13 & 4.0 & 0.4 \\
SDSSJ094818.6+023004 & 0.0604 & 2.79 & 0.14 & 2.7 & 0.5 \\
SDSSJ095842.6+631845 & 0.2426 & 22.52 & 1.75 & 21.3 & 0.8 \\
SDSSJ101342.7+125135 & 0.1391 & 1.24 & 0.13 & \nodata & \nodata \\
SDSSJ132542.3+325503 & 0.2926 & 1.71 & 0.12 & \nodata & \nodata \\
SDSSJ154322.4+331018\footnotemark[5] & 0.1265 & 3.98 & 0.15 & 5.2 & 0.4 \\
SDSSJ160808.7+394755\footnotemark[6] & 0.1909 & 12.93 & 0.14 & 35.8 & 1.5 \\
SDSSJ161910.4+064223\footnotemark[6] & 0.2100 & 45.20 & 0.14 & 77.2 & 3.0 \\
SDSSJ165958.0+213640 & 0.1567 & 2.92 & 0.14 & 2.7 & 0.5 \\
SDSSJ170859.2+322053\footnotemark[7] & 0.1206 & 1.10 & 0.14 & 2.8 & 0.5
\end{tabular}
\footnotetext[1]{The integrated flux density at 1.4 GHz is extracted from the FIRST catalogue \citep{becker95}.}
\footnotetext[2]{The RMS noise is measured locally at the source position in the FIRST catalogue.}
\footnotetext[3]{The integrated 1.4 GHz flux density is given in the NVSS catalogue \citep{condon98}. 
A typical detection limit is about 2.5 mJy in the NVSS catalogue.}
\footnotetext[4]{The mean error of the NVSS flux density.}
\footnotetext[5]{The match is uncertain for this object.}
\footnotetext[6]{In the FIRST images, a pair of radio sources is seen around the SDSS galaxy. For these objects, we use the NVSS measurements.}
\footnotetext[7]{This galaxy has a companion galaxy \citep{yamauchi08}.}
\end{minipage}
\end{table*}


The effects of radio AGN on star formation have not been investigated fully 
in post-starburst galaxies. The expected lifetime of radio emission is between $1.5~
\times ~10^{7}$ years and $10^{8}$ years \citep{alexander87,blundell00,shabala08,bird08},  
much shorter than the ages of A-type stars, 
$\sim$ 1 Gyr, found in post-starburst galaxies. 
If there is 
a close link between launching radio-mode AGN activity and quenching star formation quickly in 
the transition to the post-starburst phase, radio-emitting 
AGN activity in post-starburst galaxies should therefore be rare.


In this paper, we focus on radio-mode AGN feedback and its connection 
to quenching star formation by studying the radio properties of SDSS 
(Sloan Digital Sky Survey) \citep{york00} post-starburst galaxies 
\citep{goto07}. To this end, 
we investigate what kind of post-starburst galaxies show radio emission from AGN activity. 
Based on stellar population fits to the SDSS spectra, 
we also examine how the quantity of recent star formation 
is related to the strength of the radio-mode AGN activity. Connections between stellar populations 
and radio emission will give us insights about triggering mechanisms of recent starbursts and 
AGN activity. 

The paper is organised as follows. In \S2, we introduce our sample of post-starburst 
galaxies, and 
explain how we identify radio sources among post-starburst galaxies. 
Analysis of the stellar populations in our sample galaxies 
is given in \S3. We present the radio 
properties of our sample, as well as the constraints on mechanical feedback, in \S4. 
Discussion and conclusions are in \S5. Throughout the paper, 
we use ${\rm H_{0}~=~70~km~s^{-1}~Mpc^{-1},~\Omega_{m}~=~0.3}$, and 
${\rm{\Omega_{\Lambda}~=~0.7}}$.

\begin{figure*}
\includegraphics[scale=1.0]{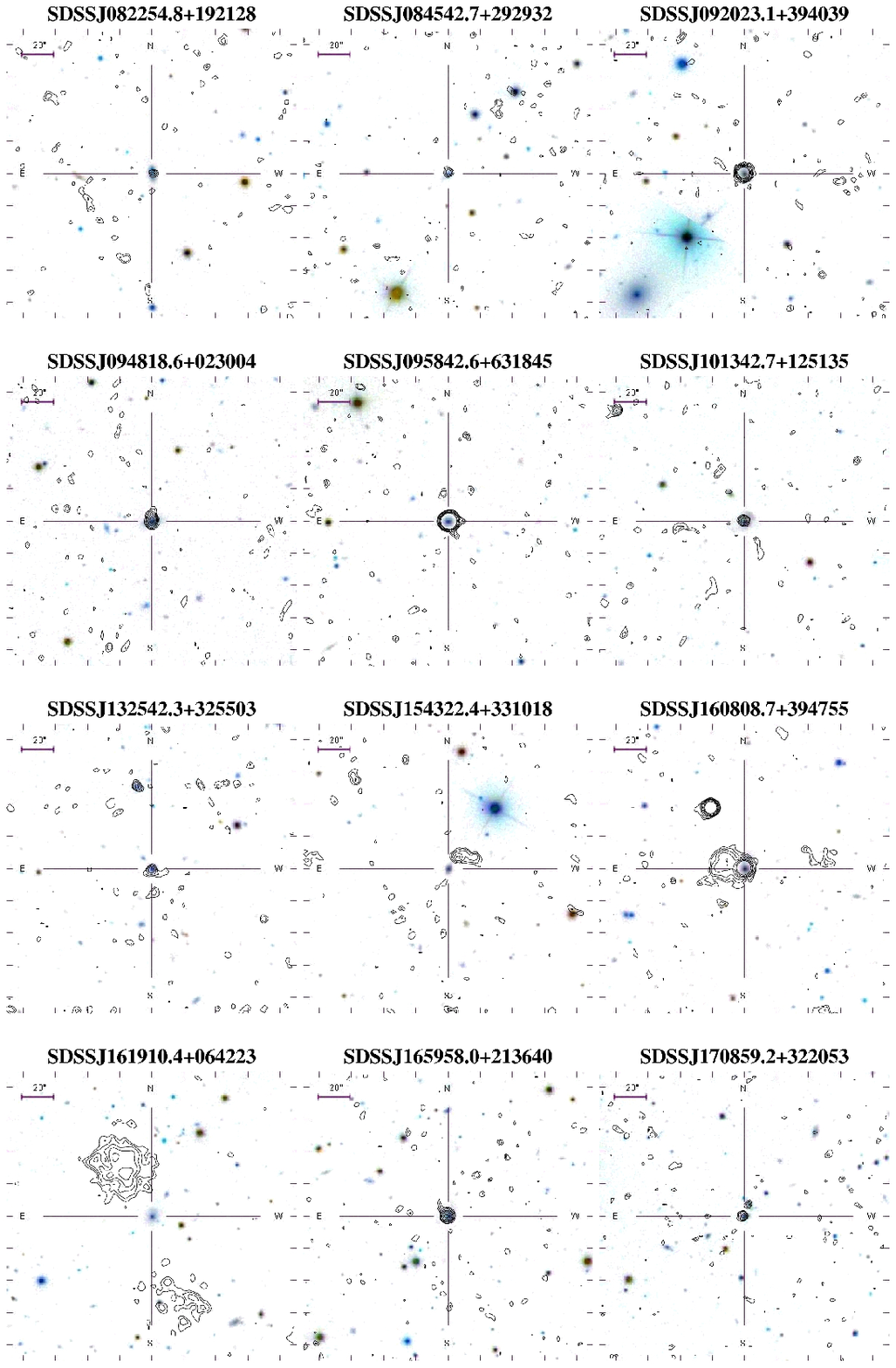}
\caption{Radio-emitting post-starburst galaxies. 
Except for SDSSJ154322.4+331018, all galaxies are well matched to the coordinates 
of radio sources. SDSSJ160808.7+394755 and SDSSJ161910.4+064223 are suspected 
to have double radio lobes. 
Colour images are the SDSS $gri$ cutout images of size 
3\arcmin$\times$3\arcmin, while the 
contour lines are from the FIRST images. \label{fig:match}}
\end{figure*}

\section{Identification of radio sources \label{sec:indentify}}

Our sample of post-starburst galaxies is from \citet{goto07}, and 
consists of galaxies that have continuum dominated by A-type stars, but are 
optically completely quiet in both star formation and AGN activity, 
as measured by the absence of both the [O II]$\lambda3727$ and 
H$\alpha$ emission lines. The sample of 
post-starburst galaxies includes all objects which do not have 
detectable [O II]$\lambda3727$ and H$\alpha$ 
emission lines in their SDSS spectra from the SDSS Data Release 5 \citep{dr5}, 
but show a rest-frame H$\delta$ equivalent width in absorption 
of $> 5$ \AA\ \citep[see][for various 
different criteria for defining E+A galaxies]{oemler09,poggianti09,falkenberg09a}. 
The redshifts of the galaxies in the sample range 
from 0.0327 to 0.3421, where the upper limit guarantees that the 
H$\alpha$ line is included in the SDSS spectra. 

We identify radio sources by matching the post-starburst galaxy sample of 564 galaxies 
to the FIRST (Faint Images of the Radio Sky at Twenty-centimetres) catalogue \citep{becker95}. 
The VLA FIRST Survey covers 
9030 ${\rm deg}^{2}$ at 1.4 GHz. 
534 galaxies in our sample lie within the FIRST coverage. We drop 22 galaxies whose spectra have 
bad pixels affecting the results of the population synthesis modelling we describe below, leaving 
513 galaxies as our main sample. 137 of these galaxies have a FIRST detection within 2 arcmin. 

We visually inspect the morphology of the radio sources in the 
FIRST images, and evaluate the goodness of the positional match between the SDSS 
galaxies and the radio sources. If radio sources are positionally coincident with other SDSS 
objects in the field, we do not consider these as good matches. 
Because the angular distance between 
the post-starburst galaxies and the other SDSS sources is generally 
more than 10\arcsec\ in these cases, 
we are unlikely to miss radio sources actually corresponding to the post-starburst galaxies. 
We find eleven unambiguous radio-emitting post-starburst galaxies, 
as well as one insecure match in which the radio source is not coincident with either the 
post-starburst galaxy or any other SDSS object, but is adjacent to the post-starburst galaxy. 
These sources are listed in Table \ref{tab:list} and Figure \ref{fig:match}. 
We conclude that the radio flux of the remaining 501 galaxies at 1.4 GHz is below the FIRST 
detection limit of about 1 mJy.

The flux limit of 1 mJy corresponds to a 
radio luminosity ranging from $10^{21.4}$ to $10^{23.7} {\rm W Hz^{-1}}$ 
for the redshifts of our sample, assuming a power-law spectral energy distribution with 
a spectral index between -1 and 1. Because this luminosity is larger than that of 
some genuine radio AGN \citep[e.g.][]{best05}, there may be radio AGNs among the sources 
undetected in the FIRST survey.

\begin{figure*}
\includegraphics[scale=0.43]{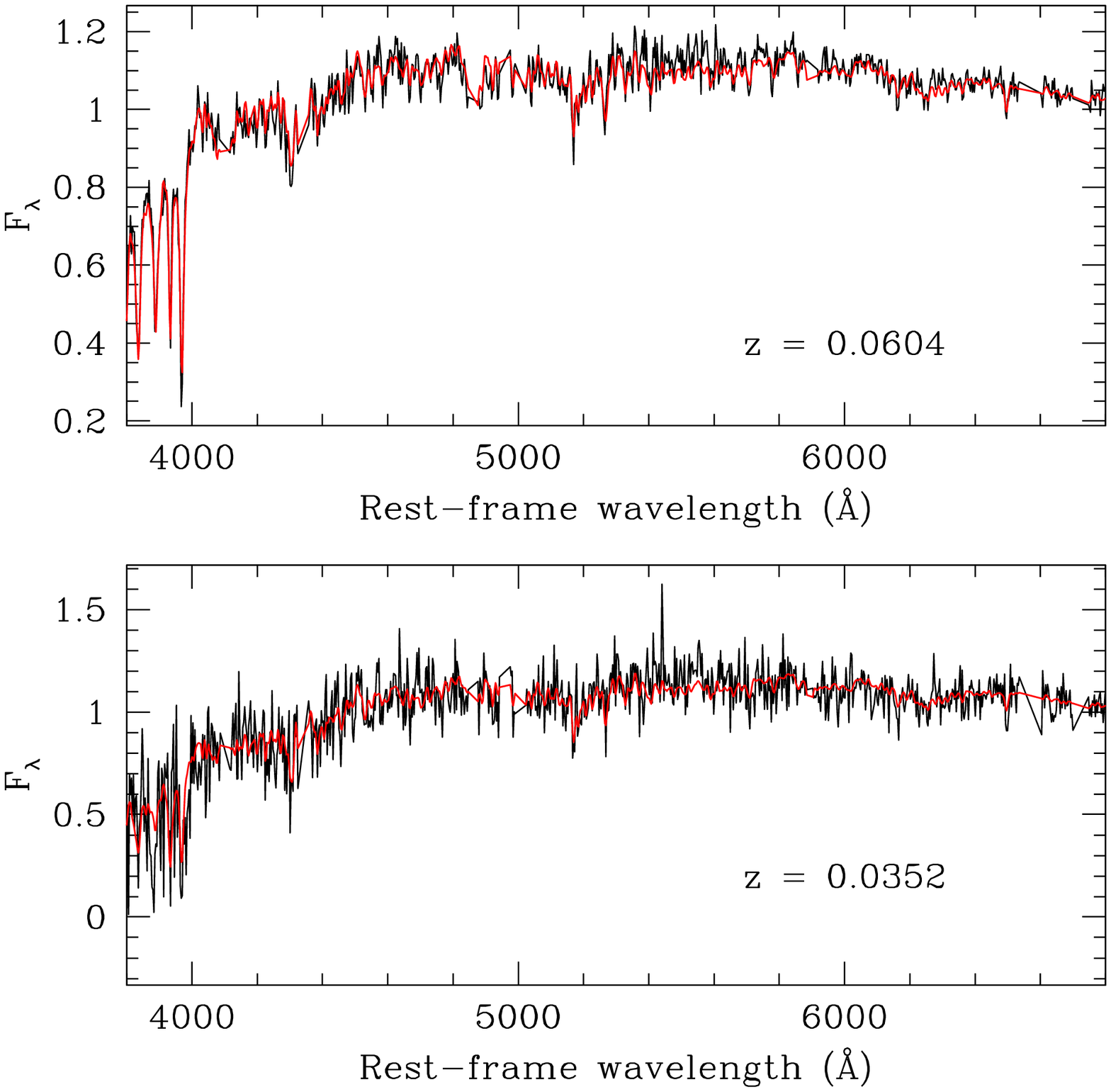}
\includegraphics[scale=0.43]{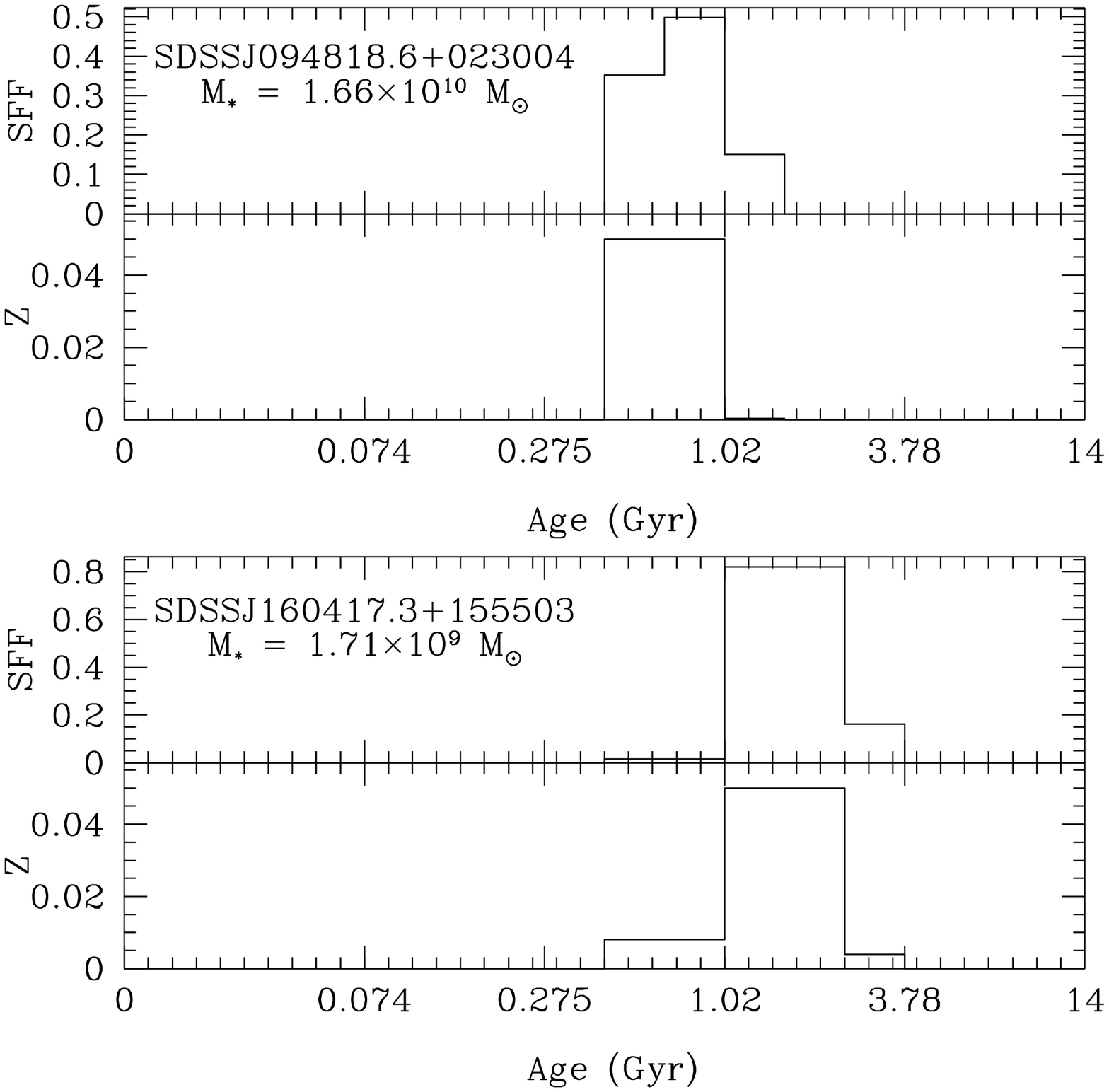}
\caption{Examples of star formation and metallicity histories with the SDSS spectra. 
SDSSJ094818.6+023004 ({\it top}) is a radio source, and 
is more massive than SDSSJ160417.3+155503, a radio-quiet source ({\it bottom}). 
The reconstructed spectra by the VESPA method are represented as red lines in the left panel.
The star formation histories (SFF; star formation fraction) and metallicity distributions (Z) 
derived by the VESPA method ({\it right}) 
show that the young stellar population is metal-rich in both spectra. 
\label{fig:exam_vespa}}
\end{figure*}

\begin{figure*}
\includegraphics[scale=0.43]{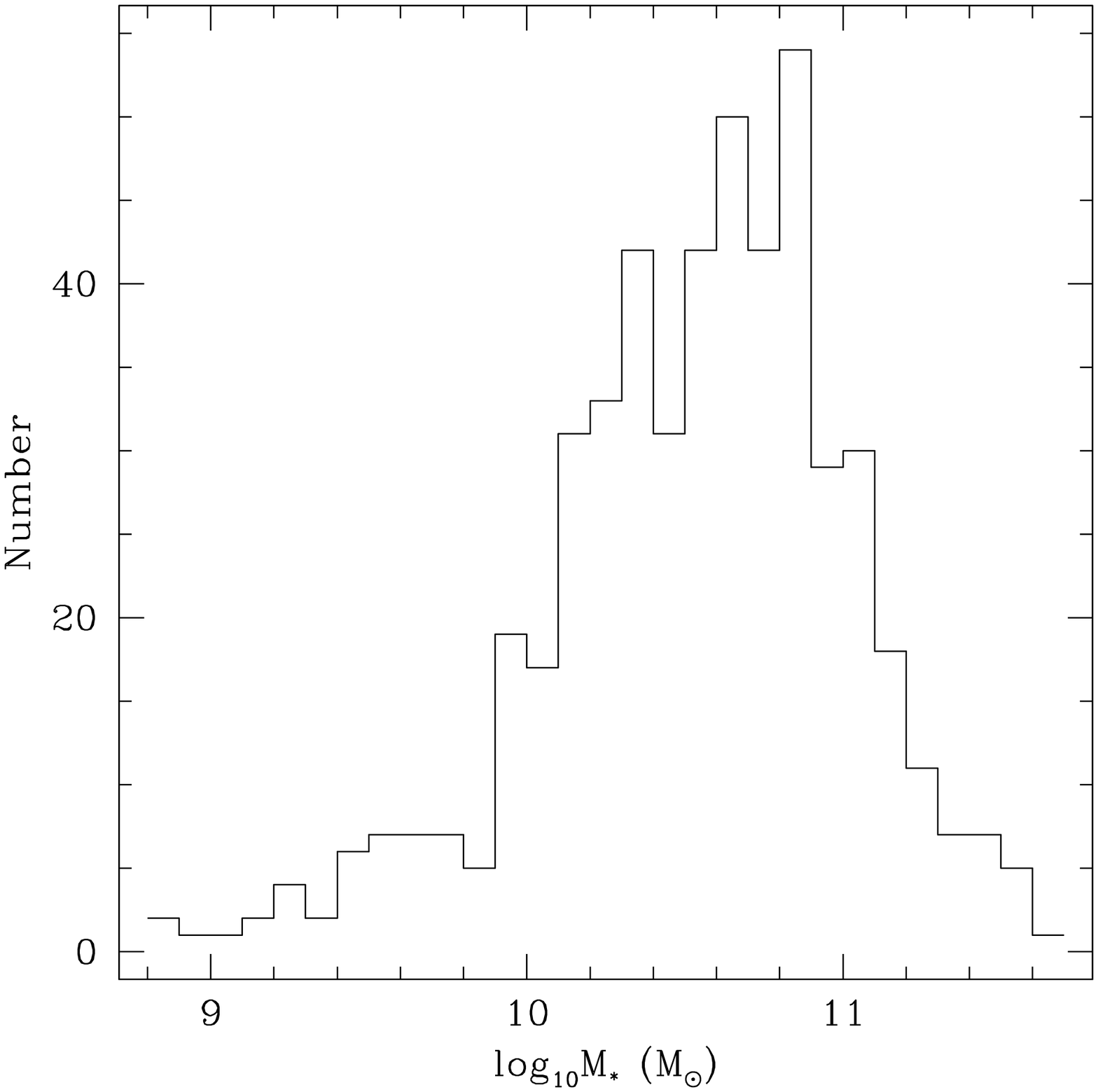}
\includegraphics[scale=0.43]{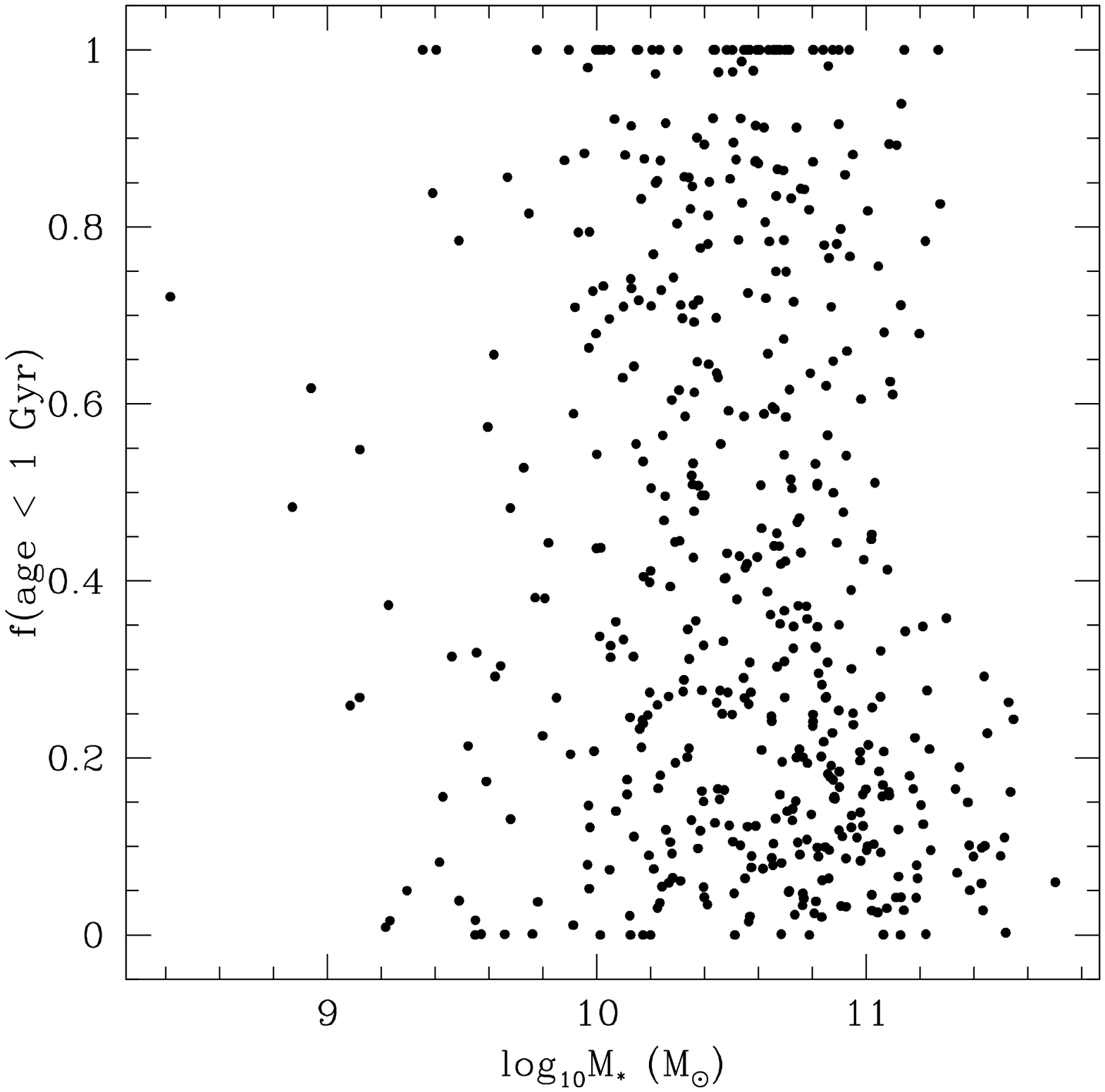}
\caption{Distribution of total stellar mass for the post-starburst galaxies in our sample 
{\it(left)}.
The mass fraction of the stellar population younger than 1 Gyr 
{\it(right)} shows a broad distribution, and no correlation with stellar mass.
\label{fig:vespa}}
\end{figure*}

Two galaxies in our sample, SDSSJ160808.7+394755 and SDSSJ161910.4+064223, have 
two resolved radio sources in the FIRST images, which we interpret as a pair of radio lobes. 
No neighbour objects around SDSSJ161910.4+064223 are matched to these sources. 
Projection of unrelated radio sources is one alternative explanation. But 
although the post-starburst galaxy does 
not lie right on the line connecting the two resolved radio sources, the redshift of the galaxy 
and the size 
of radio lobes support our inference that the host system of the lobes is the post-starburst 
galaxy. 

The positional match is less certain in the case of SDSSJ154322.4+331018. 
The bright extended radio source is offset from the galaxy by about 10\arcsec, 
and has no other optical counterpart that is positionally coincident. We assume 
that the radio source is physically associated with the post-starburst galaxy, 
but cannot prove it without further investigation of deep high-resolution 
radio imaging or deep optical imaging to find possible faint optical counterparts. 
Therefore, we tag this object as an uncertain match.

We also matched our sample against the NVSS (NRAO VLA Sky Survey) catalogue \citep{condon98}. 
The NVSS catalogue has a flux limit of 2.5 mJy, but does a better job of measuring the 
integrated flux of extended radio sources 
\citep[see][for a discussion on matching SDSS objects 
to the FIRST and NVSS catalogues]{ivezic02}. Table \ref{tab:list} compares the 
flux measurements between the FIRST and NVSS catalogue. There are serious discrepancies 
in the flux levels for the two double sources; the flux from the FIRST catalogue is 
for only one of the radio lobes 
in SDSSJ160808.7+394755 and SDSSJ161910.4+064223. When adding the fluxes from the two lobes, the 
total flux from the FIRST catalogue is comparable to the NVSS fluxes. 
Therefore, we use the flux from the NVSS catalogue for these two objects.

\begin{figure*}
\includegraphics[scale=0.43]{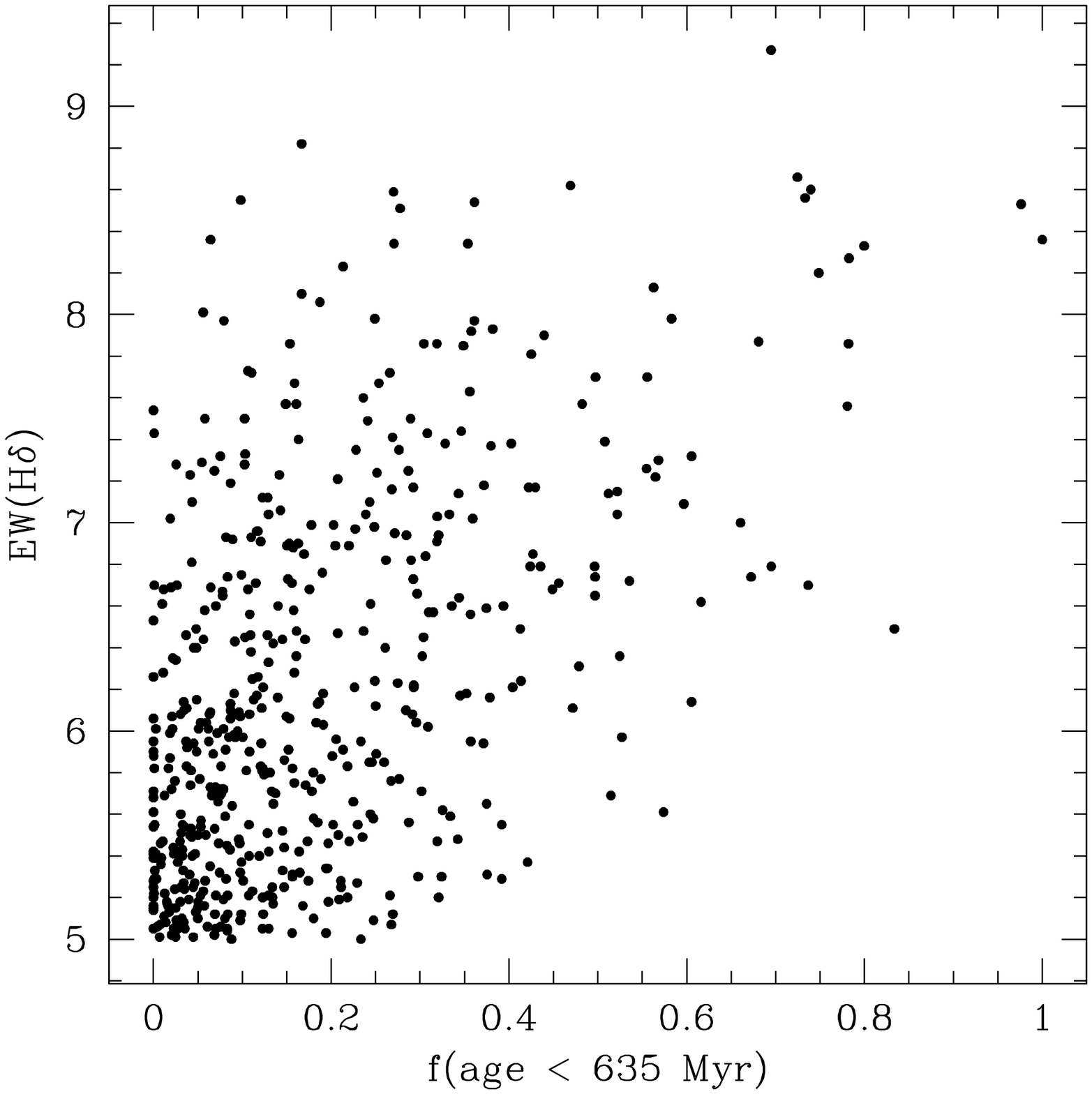}
\includegraphics[scale=0.43]{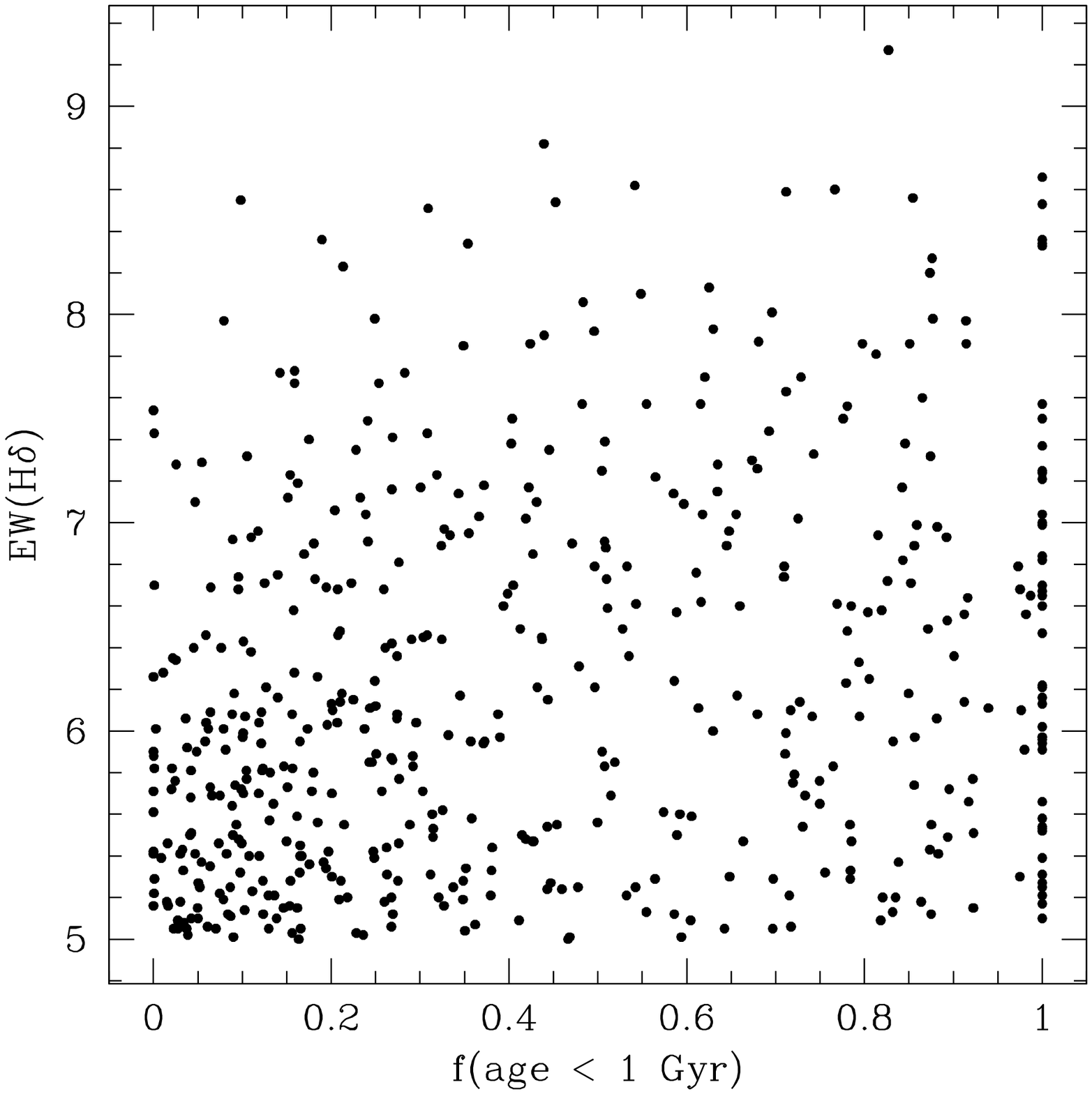}
\caption{Rest-frame equivalent width of the H$\delta$ absorption line from 
\citet{goto07} with respect to the recent star formation fraction. 
The equivalent width is weakly correlated with the fraction of stellar mass younger than 635 Myr 
({\it left}). But we do not find a correlation with the fraction of stellar mass 
younger than 1 Gyr ({\it right}) 
\citep[e.g.][]{kauffmann03b,martin07,falkenberg09a}. 
The typical error of EW(H$\delta$) is about 1\AA.
\label{fig:Hd}}
\end{figure*}

\section{Population synthesis model \label{sec:pop}}

We use the VESPA (VErsatile SPectral Analysis) method \citep{tojeiro07,tojeiro09} to determine stellar masses,  
star formation histories, and stellar metallicity distributions for all 513 galaxies in our sample. The VESPA method 
models the SDSS spectra as the linear combination of 
stellar populations in up to sixteen age bins; the metallicity of the stars formed in each bin 
can have one of five metallicities.
The number of age bins in population modelling depends on 
the amount of extractable information in the spectra given their signal-to-noise ratios. 
The model also includes one dust extinction for young stellar populations with age $< 0.3$ Gyr, 
and another dust extinction factor for old stellar populations.  

The derived star formation and metallicity histories are limited by the 3'' spatial 
coverage of the spectrograph fibres. To correct for this fibre effect, we assume that the derived 
star formation and metallicity history from the SDSS fibre spectrum are appropriate for 
the entire galaxy. Thus we scale the spectrum by the difference between the SDSS fibre magnitude 
and the Petrosian magnitude in $r$-band. The average ratio of the fibre flux to the Petrosian 
flux is about 11\% for the objects in our sample. Because we focus on the effects of 
radio-mode AGN activity which should take place at the galaxy centre, 
this fibre effect is not a serious issue in estimating the mass fraction of 
recently formed stars. 

Figure \ref{fig:exam_vespa} presents the derived star formation and metallicity 
history for two galaxies in our sample. 
SDSSJ094818.6+023004, which shows unambiguous radio emission in the FIRST data (Fig. 1), 
has H$\delta$ equivalent width (EW) ${\rm 6.18 \pm 1.02 ~ \AA}$, and most of its stellar mass is 
younger than 1 Gyr. Meanwhile, 
SDSSJ160417.3+155503, which does not have a FIRST counterpart, 
is intermediate between a strong post-starburst galaxy and 
a passively evolving galaxy, with ${\rm EW(H\delta) ~=~ 5.46 \pm 1.25 ~ \AA}$. Its 
dominant stellar population is also older than 1 Gyr. 

\begin{table*}
\begin{minipage}{126mm}
\caption{Radio luminosities and stellar masses of the identified radio sources\label{tab:radio_mass}}
\begin{tabular}{@{}lcccc} \hline \hline
SDSS name & $log_{10} L_{1.4 GHz}^{\alpha=-1}$
& $log_{10} L_{1.4 GHz}^{\alpha=0}$
& $log_{10} L_{1.4 GHz}^{\alpha=1}$
& $M_{*}$ \\ 
& (${\rm W Hz^{-1}}$) & (${\rm W Hz^{-1}}$) & (${\rm W Hz^{-1}}$) & ($10^{10} M_{\odot}$) \\ \hline
SDSSJ082254.8+192128 & 21.82 & 21.85 & 21.87 & 1.81 \\
SDSSJ084542.7+292932 & 22.75 & 22.81 & 22.87 & 2.88 \\
SDSSJ092023.1+394039 & 22.96 & 22.99 & 23.02 & 4.40 \\
SDSSJ094818.6+023004 & 22.39 & 22.41 & 22.44 & 1.66 \\
SDSSJ095842.6+631845 & 24.60 & 24.70 & 24.79 & 8.35 \\
SDSSJ101342.7+125135 & 22.81 & 22.86 & 22.92 & 4.94 \\
SDSSJ132542.3+325503 & 23.67 & 23.78 & 23.89 & 7.91 \\
SDSSJ154322.4+331018 & 23.22 & 23.28 & 23.33 & 2.20 \\
SDSSJ160808.7+394755 & 24.57 & 24.65 & 24.72 & 5.59 \\
SDSSJ161910.4+064223 & 25.00 & 25.08 & 25.16 & 9.54 \\
SDSSJ165958.0+213640 & 23.29 & 23.35 & 23.42 & 2.22 \\
SDSSJ170859.2+322053 & 22.62 & 22.67 & 22.72 & 4.88
\end{tabular}
\end{minipage}
\end{table*}

The derived stellar masses of the galaxies in the sample range from about $10^{8.4} M_{\odot}$ to about $10^{11.7} M_{\odot}$, 
as presented in Figure \ref{fig:vespa}. The typical statistical uncertainty of the stellar mass is 
about 40\% for objects in our sample, although systematic effects such as uncertainties in 
the initial mass function and the stellar population models used can affect the derived 
stellar mass \citep{tojeiro09}. Table \ref{tab:radio_mass} lists 
the stellar masses of the 12 galaxies with radio detections.

\begin{figure*}
\includegraphics[scale=0.43]{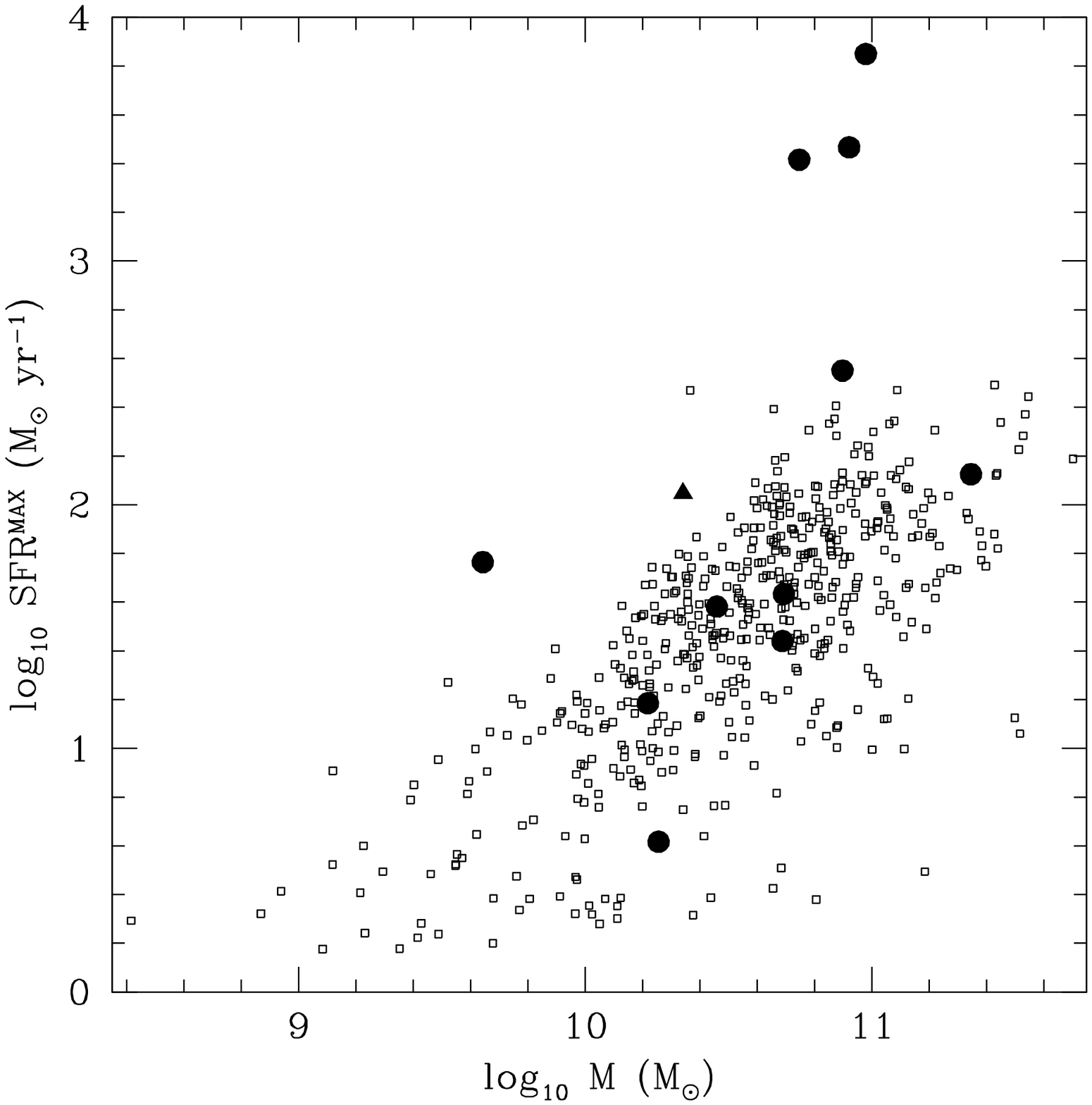}
\includegraphics[scale=0.43]{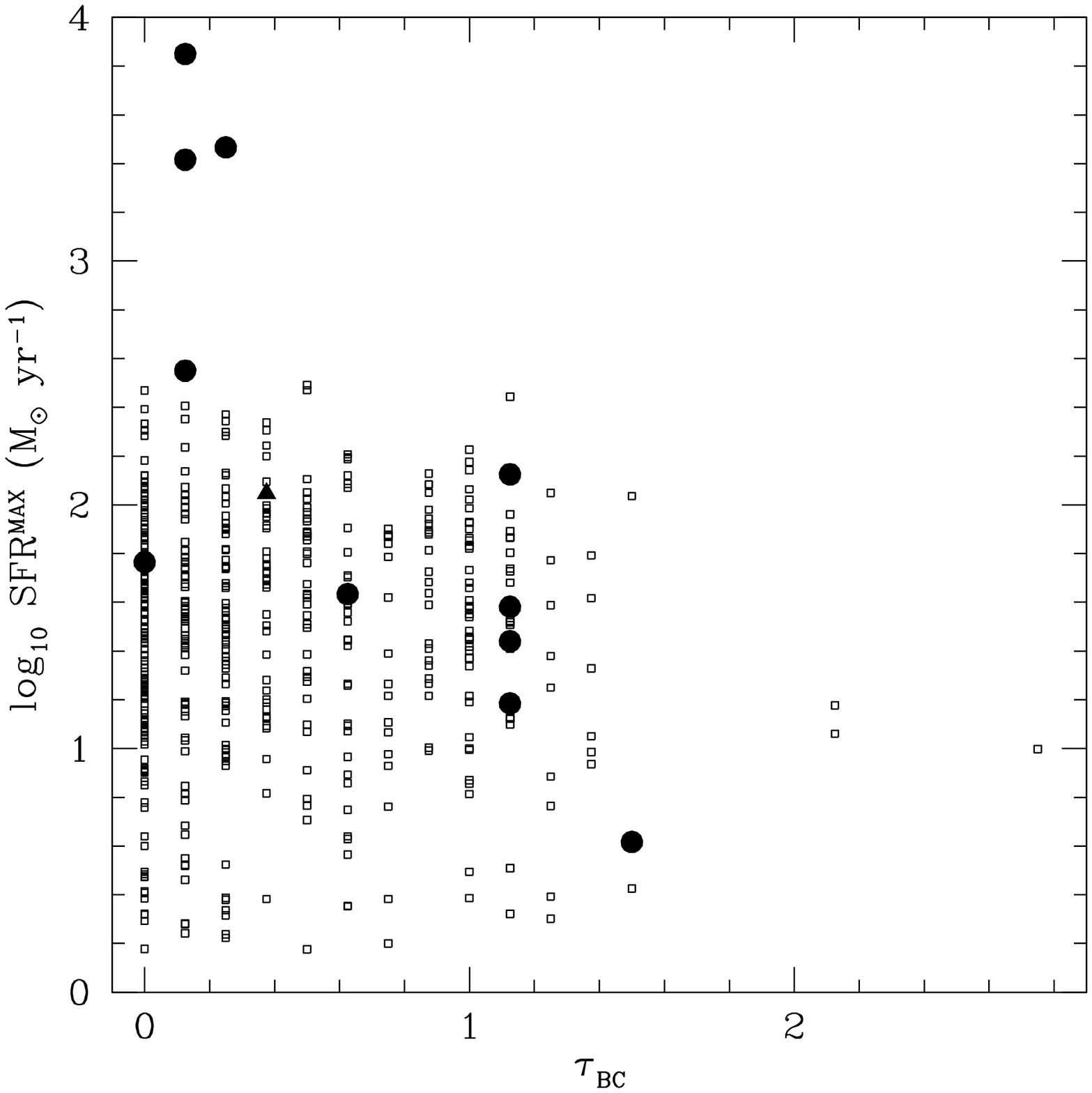}
\caption{Upper limit of the SFR derived from the observed radio emission 
(assuming a radio spectral index $\alpha = 0$) as a function of mass ({\it left}) 
and dust-extinction optical depth of the young stellar population ($\tau_{BC}$) 
from the VESPA method ({\it right}). 
For galaxies not matched to radio objects ({\it empty squares}), we plot 
the 2$\sigma$ upper limit of radio luminosity corresponding to 1 mJy flux density at 1.4 GHz. Except for 
the uncertain match SDSSJ154322.4+331018 ({\it triangle}), objects with matched radio sources are shown as 
circles. \label{fig:radio_sfr}}
\end{figure*}

An interesting quantity is the fraction of stellar mass that is recently formed in our 
sample of post-starburst galaxies. 
The fraction has a very broad distribution (see Figure \ref{fig:vespa}), 
implying that a wide range of star formation and metallicity 
histories can give rise to the spectroscopic features of post-starburst galaxies 
\citep[e.g.][]{balogh99,kauffmann03b,martin07,falkenberg09a,falkenberg09b}. 

Figure \ref{fig:Hd} shows that the mass fraction of 
young stars in galaxies is only weakly related to the strong H$\delta$ absorption line which was 
used to compile the sample of post-starburst galaxies. We used both 
635 Myr and 1 Gyr as the possible upper age limit to define the young stellar population, but we 
did not find a strong correlation between the young mass fraction and 
the strength of the H$\delta$ absorption line in either case. 
Because various star formation histories can produce a given H$\delta$ absorption line strength 
by varying the strength of recent starbursts and their 
timescales 
\citep[e.g.][]{falkenberg09a}, it is not surprising that we found no strong correlation.

\begin{figure*}
\includegraphics[scale=0.43]{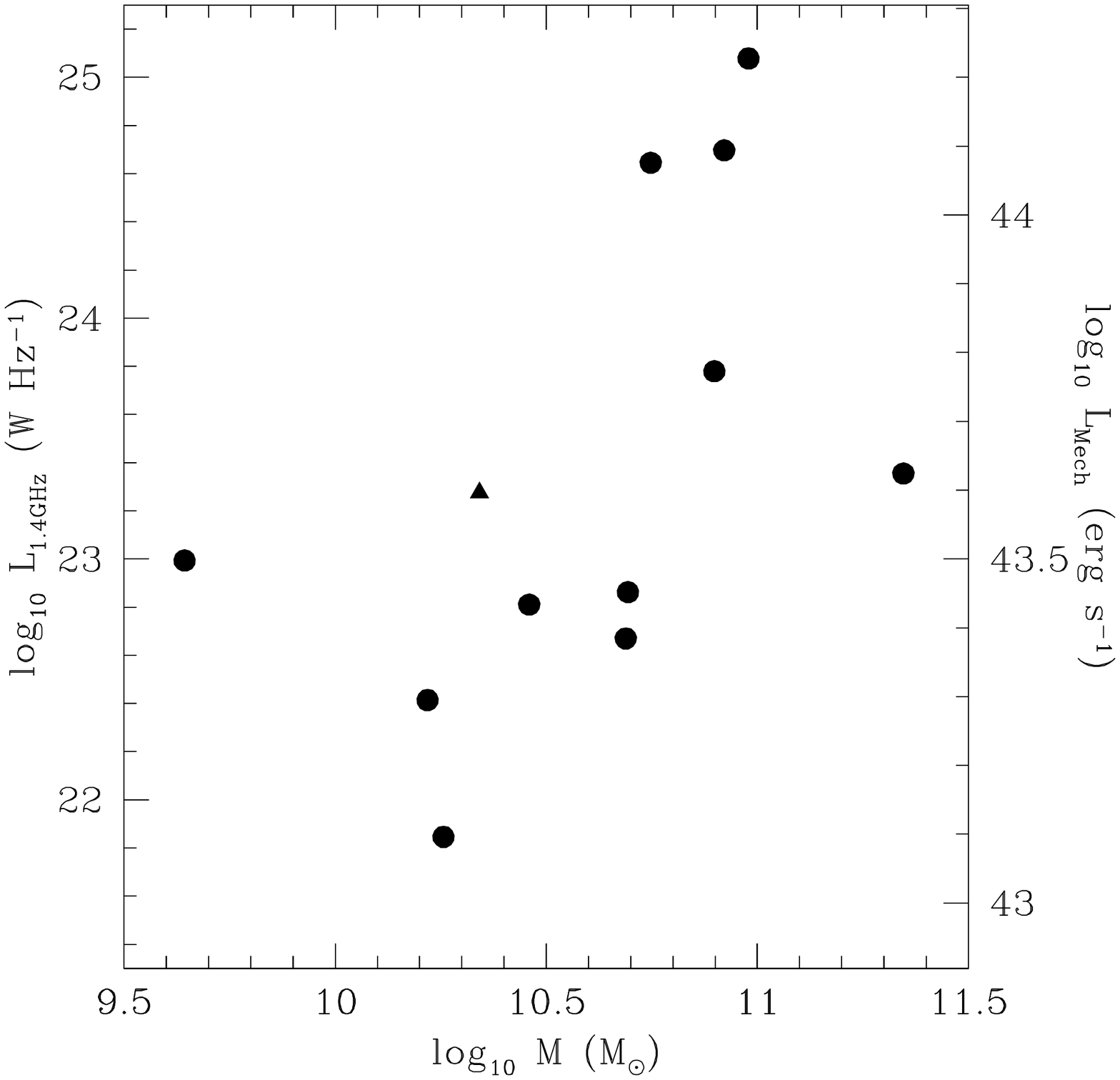}
\includegraphics[scale=0.43]{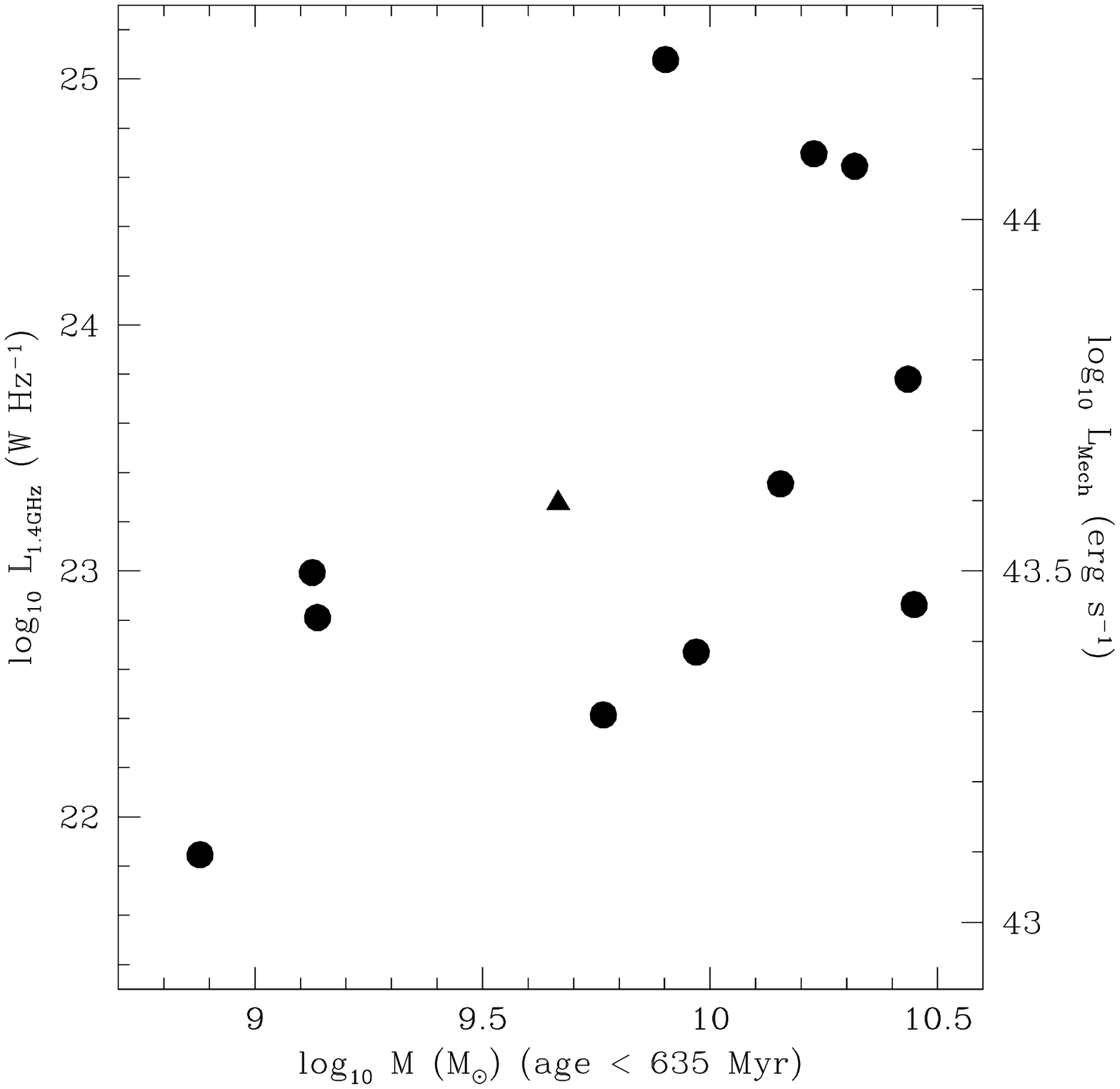}
\caption{Correlations between radio luminosity (i.e. mechanical power) and stellar mass ({\it left}), and 
the mass of the stellar population younger than 635 Myr ({\it right}). 
Symbols are same as in Figure \ref{fig:radio_sfr}. 
Among the twelve radio-detected objects, only SDSSJ101342.7+125135 has more 
than 50\% of its stellar mass younger than 635 Myr; it has a total stellar mass of 4.9 $\times 10^{10} {\rm M_{\odot}}$. 
\label{fig:radio_mass}}
\end{figure*}

\section{Radio properties} \label{sec:radio}

\subsection{Radio emission}

Radio emission in a galaxy can be due to either AGN or star formation. 
Even though none of our sample galaxies show any strong star formation signature 
in their optical spectra (by definition), 
it is possible that the galaxies are heavily dust-extincted \citep{shioya00}, or 
that weak residual star formation activity exists associated with the detected 
radio emission, which is not strong enough to produce optical emission lines. 
The timescale of radio emission from star formation can range from about 
$10^{7}$ to $10^{8}$ yr, depending on the supply and escape of cosmic rays and their 
environment as well as the lifetime of HII regions \citep{chi90,helou93}. This timescale 
might be long enough to be an explanation for at least some post-starburst galaxies. 
Therefore, it is necessary to examine whether the radio emission is due to 
star formation.

AGN radio emission can be distinguished in two ways \citep[see][for a review]{mushotzky04}. 
First, multiple radio emission components such as lobes or jets are 
signatures of AGN activity \citep{fanaroff74,reviglio06}. 
In our sample, we 
find extended multiple emission regions that are not matched to the optical light distribution 
in SDSSJ094818.6+023004, SDSSJ16080808.7+394755, and SDSSJ161910.4+064223 
as well as the uncertain case SDSSJ154322.4+331018 (see Figure \ref{fig:match}). 
Second, the radio luminosities of these galaxies can be used to identify radio-emitting AGNs. 
The radio-to-optical flux ratio or radio luminosity itself can be used statistically 
to recognise radio emission 
from AGN activity \citep[e.g.][]{ivezic02,best05}. 
But with the small number of objects in our sample, 
it is not safe to use the radio-to-optical flux ratio to distinguish radio emission 
from AGN and from star formation. Moreover, the boundary of the flux ratio 
between the two emission sources is not sharp \citep[e.g.][]{reviglio06}. 

If we assume that the radio emission is due entirely to ongoing star formation \citep[e.g.][]{goto04}, 
we can convert the observed radio luminosity to the 
expected current star formation rate (SFR) following 
the conversion given by \citet{yun01}:
\[
{\rm SFR} ~ (M_{\odot} ~ {\rm yr^{-1}}) ~ = ~ 5.9 \times 10^{-22} ~ L_{1.4 GHz} ~ ({\rm W Hz^{-1}})
\label{eq:sfr_radio}
\]
assuming the Salpeter initial mass function \citep{salpeter55} with a mass range from 0.1 to 100 ${\rm M_{\odot}}$ 
\citep[see][for different conversions]{bell03,hopkins03}. The coefficient in this equation 
has a statistical scatter of about 30\% \citep{yun01}. When we adopt the Chabrier \citep{chabrier03} initial mass function, 
the difference in the estimated SFRs with these two initial mass functions is 
${\rm log_{10} SFR_{Salpeter} - log_{10} SFR_{Chabrier}} \sim 0.186$ \citep{bardelli09}. 
For our twelve radio post-starburst galaxies, radio luminosities are estimated for three different 
spectral indices ($\alpha = -1, 0, 1$)\footnote{$\alpha = -0.5$ is the conventional 
border between steep and flat-spectrum sources.} \citep{tongue95} 
for a power-law energy distribution $F_{\nu} \propto \nu^{\alpha}$. 
The radio luminosity 
$L_{1.4 GHz}$ ranges from about $10^{21.8} {\rm W ~ Hz^{-1}}$ to $10^{25.2} {\rm W ~ Hz^{-1}}$,  
as presented in Table \ref{tab:radio_mass}. 
For the galaxies which have no detectable radio counterparts, 
we use the radio flux upper limit of 1 mJy at 1.4 GHz.

Figure \ref{fig:radio_sfr} shows the resulting predicted SFRs as a function of stellar mass. 
All galaxies with detected radio emission except SDSSJ082254.8+192128 
would have SFR $> 10 M_{\odot} ~ {\rm yr^{-1}}$ assuming the radio 
emission is due entirely to ongoing star formation. This level of hidden star formation seems highly 
unlikely given the absence of strong emission lines such as [OII] and H$\alpha$, as we now show.

We estimate the expected unextincted luminosity and equivalent width of the [OII]$\lambda$3727 emission line from 
the SFR upper limit derived earlier, assuming that the radio emission is from current star formation. 
Using the conversion from SFR to [OII] luminosity from \citet{hopkins03}, 
we derive the expected equivalent width
\[
{\rm EW}_{\rm line} ~ \sim ~ \frac{L_{\rm [OII]}}{P_{c}({\rm [OII]})} ~=~ 
\frac{SFR ~ (M_{\odot} ~ {\rm yr^{-1}}) ~ 2.97 \times 10^{33} ~ {\rm (W)}}{P_{c}({\rm [OII]})}, 
\]
where $P_{c}$ is the continuum at 3727${\rm \AA}$. 
For SDSSJ082254.8+192128, which has the lowest expected SFR $\sim 4 M_{\odot} ~ {\rm yr^{-1}}$ 
among the radio sources, we find an equivalent width of about 260 \AA , which is 
much larger than the observational limit of EW([OII]) $=$ 0.63 \AA\ \citep{goto07}. 
This constraint on the equivalent width is not affected by dust extinction, which affects 
continuum and line emission equally. It is a reasonable assumption that both emission lines and the 
blue stellar continuum originate from the same stellar population in the case of the highly obscured 
intensive star forming galaxies which we consider here. 
It is unlikely that selective high dust extinction of the emission line flux can explain the 
absence of emission lines with this large expected equivalent width.

We also examine the optical depth for dust extinction 
($\tau_{BC}$) 
which is applied to the young stellar population in the VESPA analysis \citep{tojeiro07}. 
Although $\tau_{BC}$ might be less reliable than the constraint with ${\rm EW}_{\rm line}$, 
the distribution of $\tau_{BC}$ should  be at least consistent with the constraint 
from the expected EW. Figure \ref{fig:radio_sfr} shows that $\tau_{BC}$ is low for 
sources of high radio luminosity. This trend is opposite to what would be needed to explain 
the absence of [OII] emission if the radio emission is due to star formation. We thus 
conclude that the radio emission in these twelve galaxies is dominated by AGN activity.

\subsection{Feedback energy and recent star formation}

Radio luminosity has been commonly used as a tracer of mechanical energy input by radio-mode AGN feedback. 
A simple scaling relation, albeit with non-negligible scatter, between radio luminosity and mechanical power has been suggested 
\citep[e.g.][]{birzan04}. 
Adopting the scaling relationship from \citet{birzan08}
\[
{\rm log_{10}} (\frac{L_{mech}}{10^{42} ~ {\rm erg ~ s^{-1}}}) = 0.35 {\rm log_{10}} (\frac{L_{1.4 GHz}}{10^{24} ~ {\rm W ~ Hz^{-1}}}) + 1.85,
\]
we estimate the mechanical power $L_{mech}$ from radio-mode AGN activity from 
the measured $L_{1.4 GHz}$ for the twelve objects which are matched to radio sources.

The mechanical power in radio-emitting samples ranges from $1.2 \times 10^{43}$ to 
$1.8 \times 10^{44} ~ {\rm erg / s}$, as shown in Figure \ref{fig:radio_mass}. Interestingly, 
the mechanical energy is higher for objects of higher galaxy stellar mass, 
in agreement with \citet{best05}, who show that 
luminous radio sources are more likely to be hosted by more massive galaxies. 
Although our sample is radio flux-limited, 
the absence of luminous radio sources for low-mass galaxies is not due to selection effects, 
as the distribution of the radio luminosity upper limits for objects 
without radio counterparts shows in Figure \ref{fig:radio_sfr}. 
This trend is not surprising because massive galaxies host massive central black holes, 
which can produce radio jets with a large amount of mechanical power \citep[see][for a discussion]
{meier03,best05,bardelli10}. It also suggests that the mechanical power is correlated with the 
old stellar population, which dominates the stellar mass.

In contrast, no correlation of recently formed stellar mass with mechanical 
power output is apparent, as presented in Figure \ref{fig:radio_mass}. In elliptical 
galaxies with hot gas halos, \citet{allen06} found a tight positive correlation between 
Bondi accretion power and jet power. Because the SFR also shows a strong correlation with the 
black hole growth rate at the centre of cooling flows \citep{rafferty06}, a correlation 
between jet power and recently formed stellar mass is expected 
in galaxies with significant cooling flows. But our sample is not consistent with this 
picture. This result suggests that the recent star formation was ignited by a cooling flow 
in only a fraction at best of radio post-starburst galaxies in our sample.

\section{Discussion and conclusions}

We find that the abrupt cessation of star formation in post-starburst galaxies 
is not caused solely by radio-mode feedback from AGN. 
The typical lifetime of radio emission is of 
order $10^{7}$ - $10^{8}$ yrs \citep{blundell00,shabala08,bird08}. 
However, the post-starburst phase can last 
up to $\sim$ 1 Gyr (the lifetime of A-type stars), and must be longer than $\sim$ 
1 Myr which is constrained by the typical lifetime of massive stars that produce 
ionising photons and emission lines \citep{charlot00}. Therefore, the existence of 
radio-emitting post-starburst galaxies implies that radio-mode AGN activity, which 
is the source of the radio emission we observe, should form after the end of the recent starburst. 

But we cannot rule out that the rest of the objects in our sample 
(i.e. those without detectable radio emission) might have experienced 
radio-mode feedback and the cessation of star formation concurrently. 
Because our sample is flux-limited in the radio, we were unable to prove that the late stage of 
post-starburst galaxies always accompanies radio-mode AGN activity after the end of 
star formation, particularly for massive post-starburst galaxies. Future deep radio surveys 
of post-starburst galaxies will be required to detect low-luminosity radio emission 
in the rest of our sample galaxies.

This time delay between the cessation of star formation and radio-emitting AGN activity 
was already suggested for some radio-excess IRAS galaxies based on the same kind 
of time constraints arguments as we have given \citep{buchanan06}. 
This kind of time delay is also found in the nuclei of local Seyfert galaxies based on ages of stellar 
populations \citep{davies07}. 
Although radio excess in IRAS galaxies and Seyfert galaxies might evolved differently than 
radio emission in post-starburst galaxies, the delay between fuelling 
the central black hole and recent star formation is about 50 to a few hundred Myr \citep[e.g.][]{schawinski07}. 
In our constraint, the upper limit 
of this time delay can be up to about 1 Gyr, which is basically limited by the age of post-starburst 
stellar population. The physics of this time delay is still uncertain, although 
observational constraints suggest that strong fuelling onto 
a central supermassive black hole occurs after intensive star formation, and is regulated by the 
recent and ongoing star formation in a galaxy \citep[see][for a discussion of the time delay]
{shlosman90,davies07}.

In the {\it cooling 
flow paradigm} of early-type galaxies, cold gas in the cooling flow forms new stars and also 
fuels the central black hole. If so \citep{cardiel98,bildfell08,odea08,pipino09}, the 
spectral energy distribution might go through a post-starburst phase, with AGN feedback 
\citep[e.g.][]{ciotti10}. 
Radio-emitting post-starburst galaxies without evidence for galaxy interactions might have 
been fed by a cooling flow.

Alternatively, let us consider the galaxy interaction scenario. Many post-starburst galaxies are thought to be 
merger remnants, where the termination of star formation seems to occur before the radio-mode AGN activity 
\citep[e.g.][]{tadhunter05,emonts06}. 
Moreover, powerful radio galaxies seem to be more commonly 
hosted by galaxy mergers than are less luminous radio galaxies and radio-quiet ellipticals 
\citep{heckman86,wilson95}. One of our sources, SDSSJ170859.2+322053, is already known to be a 
merging/interacting galaxy \citep{yamauchi08}, but follow-up deep optical imaging 
of the other radio-emitting post-starburst galaxies in our sample may reveal mergers, 
allowing an investigation of a link between galaxy interaction, star formation 
triggering, star formation quenching, and the initiation of radio-mode AGN activity. 
Numerical simulations of galaxy mergers or interactions including 
radio-mode AGN activity would allow a test of 
the observed features of morphologically disturbed radio-emitting post-starburst galaxies 
\citep[e.g.][]{sijacki07}.

The key difference between the two scenarios might be 
different time delays between the end of 
intensive star formation and the ignition of AGN activity \citep[e.g.][]{wills08}. 
Further investigation of distinctive spectroscopic features from different 
time delays and star formation histories may yield 
the evidence of these two processes in the 
post-starburst phase \citep[e.g.][]{falkenberg09a}.

In this paper, we did not consider that possible early radio AGN activity, which does not 
correspond to what we detect now in radio, might be an effective way to quench 
star formation. For example, the radio galaxy 3C 236 has both multiple stellar populations and 
multiple radio sources which might have formed concurrently in several episodes 
\citep{tremblay10}. But because the visible radio relics of the past active phase dim fast and 
appears at low frequencies within 1 Gyr, it might be difficult to detect activity 
older than about $10^{8}$ yr in radio data \citep{kaiser02,godambe09}. 
Deep X-ray imaging of radio-emitting post-starburst galaxies might be helpful to discover 
the fossil record of the past radio-mode AGN activity \citep{juett08}.

In addition to the delayed radio-mode AGN activity discussed in this paper, 
there might be other kinds of AGN feedback effects which can cause the sudden 
cessation of star formation in post-starburst galaxies. For instance, 
it is already known that some post-starburst galaxies show clear spectroscopic features of 
quasar emission, implying that quasar activity lasts longer than star formation or is triggered after 
the peak of star formation \citep[e.g.][]{brotherton99}. X-ray emission from AGN accretion is also 
found in some post-starburst galaxies \citep[e.g.][]{brown09}. 
It will be important to understand what kind of 
post-starburst galaxies do not have any kinds of AGN feedback effects, and how the types of 
AGN activity are related to other properties of post-starburst galaxies such as merger stages and 
environment.

\section*{Acknowledgements}

We are grateful to Jeremiah Ostriker, James Gunn, 
Gillian Knapp, Renyue Cen, and Christy Tremonti 
for useful discussions and careful reading of the manuscript. 
We would like to thank Eric Pellegrini and Ryan Porter for useful discussions. 
We thank the anonymous referee for comments which improved this manuscript. 
M.-S.S. was supported by the Charlotte Elizabeth Procter Fellowship of Princeton
University. M.-S.S. and M.A.S. acknowledge the support of NSF grant AST-0707266. 

Funding for the SDSS and SDSS-II has been provided by the Alfred P. Sloan Foundation, 
the Participating Institutions, the National Science Foundation, the U.S. Department of Energy, 
the National Aeronautics and Space Administration, the Japanese Monbukagakusho, the Max Planck Society, 
and the Higher Education Funding Council for England. The SDSS Web Site is http://www.sdss.org/.
The SDSS is managed by the Astrophysical Research Consortium for the Participating Institutions. 
The Participating Institutions are the American Museum of Natural History, Astrophysical Institute Potsdam, 
University of Basel, University of Cambridge, Case Western Reserve University, University of Chicago, 
Drexel University, Fermilab, the Institute for Advanced Study, the Japan Participation Group, 
Johns Hopkins University, the Joint Institute for Nuclear Astrophysics, the Kavli Institute for Particle 
Astrophysics and Cosmology, the Korean Scientist Group, the Chinese Academy of Sciences (LAMOST), 
Los Alamos National Laboratory, the Max-Planck-Institute for Astronomy (MPIA), the Max-Planck-Institute 
for Astrophysics (MPA), New Mexico State University, Ohio State University, University of Pittsburgh, 
University of Portsmouth, Princeton University, the United States Naval Observatory, and the University of Washington.

\end{document}